\documentclass[journal]{IEEEtran}

\ifCLASSINFOpdf
\else
\fi
\usepackage[hyphens]{url}
\usepackage{graphicx}
\usepackage{caption}
\usepackage{tabularx,ragged2e}
\usepackage{booktabs}
\usepackage{multirow}
\usepackage{siunitx}
\usepackage{pifont}
\newcommand{\cmark}{\ding{51}}%
\newcommand{\xmark}{\ding{55}}%
\usepackage{amssymb}
\newcolumntype{C}{>{\Centering\arraybackslash\hspace{0pt}}X}
\usepackage[table]{xcolor}
\usepackage{threeparttable, tablefootnote}

\begin{document}
%
\title{WavRx: a Disease-Agnostic, Generalizable, and Privacy-Preserving Speech Health Diagnostic Model}
%
%
%

\author{Yi Zhu,~\IEEEmembership{Graduate Student Member,~IEEE,}
        and Tiago Falk,~\IEEEmembership{Senior Member,~IEEE,}

\thanks{Authors are with INRS. E-mail: Yi.Zhu@inrs.ca}
\thanks{Manuscript received XXX.}}

%
%

\markboth{}%
{Shell \MakeLowercase{\textit{et al.}}: Bare Demo of IEEEtran.cls for IEEE Journals}
%



\maketitle

\begin{abstract}
Speech is known to carry health-related attributes, which has emerged as a novel venue for remote and long-term health monitoring. However, existing models are usually tailored for a specific type of disease, and have been shown to lack generalizability across datasets. Furthermore, concerns have been raised recently towards the leakage of speaker identity from health embeddings. To mitigate these limitations, we propose WavRx, a speech health diagnostics model that captures the respiration and articulation related dynamics from a universal speech representation. Our in-domain and cross-domain experiments on six pathological speech datasets demonstrate WavRx as a new state-of-the-art health diagnostic model. Furthermore, we show that the amount of speaker identity entailed in the WavRx health embeddings is significantly reduced without extra guidance during training. An in-depth analysis of the model was performed, thus providing physiological interpretation of its improved generalizability and privacy-preserving ability.
\end{abstract}


%
\IEEEpeerreviewmaketitle

\section{Introduction}
\IEEEPARstart{D}{uring} recent years, speech has emerged as a promising modality for disease diagnosis and remote health monitoring. Speech health diagnostics is typically based on the assumption that diseases causing abnormalities in articulatory and/or respiratory systems would lead to an atypical pattern in the human voice signal~\cite{fagherazzi2021voice}. Such abnormality could be due to a variety of reasons, such as impaired neuromuscular control or an inflammation in the vocal tract and lungs~\cite{fagherazzi2021voice}. While the impact on the speech signal may sometimes be imperceptible to humans, a machine learning (ML) model could be trained to detect certain disease-related vocal biomarkers.

Over the years, there has been a substantial body of work that has explored the use of speech processing for diagnostics, including but not limited to COVID-19~\cite{deshpande2020overview}, dysarthria~\cite{millet2019learning}, Parkinson's~\cite{gullapalli2022early} and Alzheimer's disease~\cite{konig2015automatic}, as well as many other general respiratory symptoms~\cite{xia2021covid}. Several challenges related to speech-based health diagnostics have also emerged, showing the impact of deep learning~\cite{cummins2018speech}.
Despite these many published works, very few systems exist commercially or are used in real-world settings. There may be several reasons for this. First, existing system architectures are usually tailored for a single type of disease, i.e., are \emph{disease-dependent}. While disease-related biomarkers can be well captured by the models, other health attributes are likely to be overlooked. For example, systems that focus on speech intelligibility may be useful at diagnosing dysarthria, but may fail at detecting COVID-19 infections. As recent innovations in self-supervised learning (SSL) are showing~\cite{hsu2021hubert, chen2022wavlm, baevski2020wav2vec}, it is possible to learn universal representations that can be used across many different downstream tasks~\cite{yang2021superb}. The same is desirable for health diagnostics, where  the same architecture can be applied to a variety of diseases (i.e., \emph{disease-agnostic}) where only downstream fine-tuning is needed. 

Second, it is expected that a well-trained diagnostic model will generalize well across datasets that share the same or similar pathology. However, recent studies have reported severe degradation in performance across several state-of-the-art diagnostic systems when tested on unseen data from different patients with the same disease~\cite{zhu2022generalizable}. This has been attributed to different confounding factors (e.g., noise level, gender) generated unwarily during data collection~\cite{zhu2023investigating, schu2023using}. These factors could lead to models, especially deep learning-based ones, to overfit to a certain database property (e.g., changes in sampling frequency for different disease labels \cite{coppock2023summary}) and not necessarily to diagnostic information. The lack of generalizability makes the reliability of existing models questionable and further exacerbates the criticism around the lack of explainability and the ``black-box'' aspect of deep neural networks. 

Lastly, since voice carries personal identity attributes, such as gender, age, and race~\cite{schuller2013paralinguistics}, uploading the voice signal to an online platform for model training and evaluation is dangerous, especially considering the rapidly growing voice cloning techniques~\cite{dagar2022literature}. One method to alleviate this privacy concern is to extract a speech representation locally, then upload only the representation itself. However, studies have shown that health information is likely to entangle with speaker identity in most widely used speech representations (e.g., openSMILE features, ECAPA-TDNN embeddings, and universal representations)~\cite{zhu2024impact, javanmardi2023wav2vec, chen2022supervised}, suggesting that existing health representations still suffer from speaker leakage. While some privacy-preserving methods have been proposed as an alternative, including adversarial training~\cite{ravi2022step} and voice anonymization~\cite{das2023stargan}, such methods may alter the speech signal, thus potentially removing disease-discriminatory details; this was recently shown to be the case for COVID-19 detection~\cite{zhu2024impact}. 


To tackle these three major limitations, in this paper we propose a new speech health diagnostic model that is disease-agnostic, generalizable across datasets, and privacy-preserving. The proposed model, termed WavRx, is built on top of the well-known WavLM model~\cite{chen2022wavlm} and incorporates a novel modulation dynamics module, which mixes the high-resolution temporal WavLM representation with the long-term modulation dynamics of speech. While the WavLM representation can carry both linguistic and paralinguistic attributes~\cite{chen2022wavlm} at a \SI{50}{Hz} rate, these attributes focus more on transient temporal changes. Articulation and respiration related abnormalities, on the other hand, may modulate these short-time features at a much lower rate. As such, the proposed modulation dynamics block is designed to capture long-term variability and to provide complementary information to the temporal details. 

Our main contributions in this paper can be summarized as follows:
\begin{enumerate}
    \item We propose a new speech health diagnostics model, WavRx, that mixes the universal temporal representation with long-term modulation dynamics. WavRx is tested on six different pathological speech datasets, spanning four different speech pathologies, and achieves state-of-the-art (SOTA) performance on 4 out of 6, with the highest average performance among six benchmark models.
    \item We show that the modulation dynamics block, while being parameter-free, can significantly improve the overall model generalizability across datasets and diseases that share similar symptoms.
    \item We demonstrate that the modulation dynamics block helps to markedly remove the speaker attributes from the health embeddings learned by WavRx, without the need for extra guidance during training.
    \item We find that the health embeddings learned from the dynamics representation are twice more sparse than from the temporal representation, which helps to remove disease-irrelevant information.
\end{enumerate}

\section{Related work}
\subsection{Speech-based diagnostic models}
Earlier works have focused on knowledge-based features to characterize the underlying speech pathology. Besides conventional speech features, such as mel-spectrograms or mel-frequency cepstral coefficients (MFCCs), studies have examined a wide variety of features associated with health status. The openSMILE ComParE set~\cite{eyben2010opensmile}, for example, has been used as a baseline across several challenges, such as the 2021 COVID-19 detection challenge~\cite{schuller2021interspeech}, the 2017 cold\&snoring recognition challenge~\cite{schuller2017interspeech}, and the 2012 pathology sub-challenge that predicts speech intelligibility for individuals that received cervical cancer surgeries~\cite{schuller2021interspeech}. 

Other studies have proposed features designed specifically for certain types of diseases, such as phonation and articulation features for Parkinson's disease~\cite{arias2017parkinson, vasquez2018towards}, linear prediction (LP) based features for COVID-19~\cite{zhu2023covid}, and voice quality features for depression~\cite{afshan2018effectiveness}, just to name a few. These hand-crafted features aim to capture certain aspects of the speech signal affected by the disease using signal processing techniques. The engineered features are then fed into classical ML classifiers, such as support vector machine or random forest classifiers. Major advantages of hand-crated features are that they provide some explainability and interpretability (e.g., LP residuals represent vocal cord vibration patterns), are suitable for small datasets, which are typically the case in healthcare settings, and tend to generalize better across datasets \cite{zhu2022fusion}. 

More recently, models based on deep learning (DL) have started to burgeon \cite{cummins2018speech}. These models typically take  as input the speech waveform or some variant of the spectrogram (e.g., a mel-scaled spectrogram) and learn the underlying biomarkers via a data-driven approach. DL models are usually designed for one specific type of disease. For example, convolutional (CNN) and recurrent (RNN) networks have been used for COVID-19 detection using cough, speech, and breathing signals as input~\cite{coppock2021end, sharma2022second, zhu2024spectral}. These models were trained from scratch using a limited amount of data, hence their power is yet to be fully explored. To address this issue, some studies have investigated transfer learning with large-size pre-trained models, such as the VGGish networks~\cite{simonyan2014very}, ECAPA-TDNN~\cite{desplanques2020ecapa}, and audio transformers~\cite{gong2021ast}. Studies have shown that pre-training on out-of-domain data (e.g., image datasets, speaker verification datasets, audio events) could also benefit speech diagnostics performance~\cite{xia2021covid, sekhar2022dysarthric, jeancolas2021x, coppock2022audio}. 

While pre-training is usually conducted in a supervised manner, there have been some initial attempts to leverage SSL pre-trained models for diagnostics~\cite{chen2022supervised, xue2021exploring, violeta2022investigating}. The underlying assumption is that the universal speech representation resultant from the models, such as Wav2vec~\cite{baevski2020wav2vec}, carries a variety of speech information, including linguistic, paralinguistic, and diagnostic information~\cite{baevski2020wav2vec, hsu2021hubert, chen2022wavlm}. It has been shown that self-supervised pre-training is less biased by the upstream datasets than by supervised training~\cite{lu2022self}, thus making universal speech representations an excellent candidate for diagnostic tasks. 

However, while most existing works have taken different  universal representations directly as the feature input to downstream diagnostic classifiers (e.g., \cite{chen2022supervised, xue2021exploring, violeta2022investigating}), we argue that existing universal representations are suboptimal for diagnostics tasks due to two main reasons. First, SSL models, such as Wav2vec~\cite{baevski2020wav2vec}, WavLM~\cite{chen2022wavlm}, and HuBERT~\cite{hsu2021hubert}, aggregate the input waveform into short segments by a convolutional layer before feeding into the transformer layers. In the case of WavLM, the receptive field of each unit in the CNN output is around \SI{20}{ms}, similar to the frame lengths used in the mel-spectrogram. While this is short enough to capture linguistic content (e.g., phonemes) as well as other temporal details (e.g., speaker details), more longer-term dynamics, such as speaking rate, respiration, and emotions, may not be well encoded. This corroborates with the improved performance achieved by appending different downstream layers to the temporal representation, such as 1D CNNs (\cite{pepino2021emotion}) and RNNs (\cite{zhang2021depa}). Second, the existence of the linguistic content in temporal representation may bias the diagnostic models, as the disease biomarkers should be independent of spoken content. Given these limitations, we propose a new representation that also captures long-term dynamics, following the widely-used concept of the speech modulation spectrum, but instead, applied to a universal speech representation. 

\subsection{Modulation dynamics of speech}
Speech is produced by the vibration of vocal cords, the vibration is then transmitted through the vocal tract and modulated by the articulatory movement and respiration, generating the speech hearable by humans~\cite{casserly2010speech, ohala1990respiratory}. Typical speech analysis focuses on short-time analysis to capture transient changes caused by changes in phonemes. For example, the window size for the short-time Fourier transform (STFT) is usually \SI{8}{} to \SI{32}{ms}~\cite{rabiner2007introduction}. Speech modulation, in turn, changes at a much lower rate due to the limit of human physiology. For articulatory movement, for instance, between 2 and 10 syllables are being uttered per second for most of the languages~\cite{hermansky1998modulation}. However, such underlying modulation is not well captured by a spectrogram and measures such as delta and double-delta cepstral parameters have been used for decades as measures of velocity and acceleration of changes in the cepstral parameters over somewhat larger window durations.

To address this issue, several researchers have relied on the so-called modulation spectrum (e.g., \cite{greenberg1997modulation,falk2010non}), which applies a second STFT to each frequency component obtained from the spectrogram. This extends the conventional spectrogram to a 3-dimensional space with an added modulation frequency axis. With a window size of over \SI{128}{ms}, the modulation spectrum analyzes the hidden periodicity of human speech. While most of the linguistic content is lost in the modulation frequency domain, other vocal characteristics such as speaking rate~\cite{hermansky1998modulation}, vocal hoarseness~\cite{markaki2011voice}, and whispering~\cite{sarria2013whispered} may be better manifested. Features derived from such representation have been previously applied in the detection of dysarthric speech~\cite{falk2012characterization}, whispered speech~\cite{sarria2013whispered}, voice pathologies~\cite{markaki2011voice}, COVID-19~\cite{zhu2023covid}, and emotional speech~\cite{wu2011automatic}, to name a few. Motivated by the idea of the modulation spectrum, we here applied the modulation transformation to the universal representations to better capture their health-related attributes.

\section{Proposed model architecture}
The proposed WavRx comprises three main components: (1) a pre-trained encoder to extract temporal representations from the raw waveform; (2) the modulation dynamics block to capture long-term dynamics of the encoded temporal representations; and (3) attentive statistic pooling and output layers to fuse representations from the previous two blocks and generate a final decision. Details about each component are described in the following subsections. 

\begin{figure*}
    \centering
    \includegraphics[width=\linewidth]{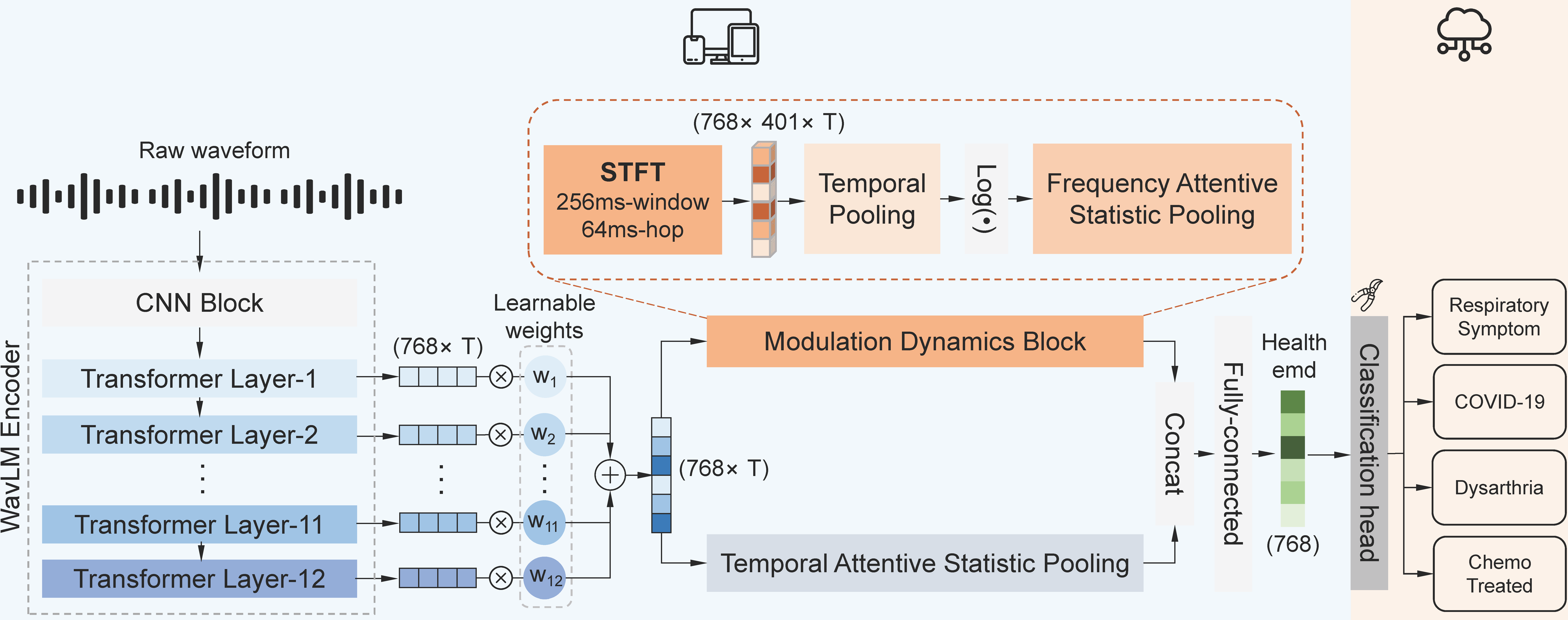}
    \caption{Architecture of the proposed WavRx model.}
    \label{fig:model}
\end{figure*}

The model architecture is depicted in Figure~\ref{fig:model}. Considering the privacy requirement in real-world applications, WavRx encoder can be deployed locally to extract health embeddings, which are then uploaded to a central cloud server for decision-making. In later sections, we show that the health embeddings entail minimal speaker identity information, hence preventing the leakage of user identity. Our code is made publicly available on GitHub\footnote{\url{https://github.com/zhu00121/WavRx}}. Owing to the data sharing terms of the employed datasets, pre-trained model backbones are released upon requests.

\subsection{Temporal representation encoder}
The proposed model builds on top of the pre-trained WavLM as the temporal representation encoder~\cite{chen2022wavlm}. WavLM takes a raw speech waveform as input and firstly feeds it into a CNN block comprised 7 temporal CNN layers with 512 channels, cascaded by layer normalization and a GELU activation layer. Each time step in the output from CNN block represents \SI{25}{ms} of audio with \SI{20}{ms} hop length. The CNN output is then sent into a transformer backbone, which comprises 13 layers with 768-dimensional hidden states. We employed the WavLM Base+ version\footnote{HuggingFace link: https://huggingface.co/microsoft/wavlm-base-plus. Accessed May 23rd, 2024.} which was pre-trained on 60K hours of Libri-light~\cite{kahn2020libri}, 10K hours of Gigaspeech~\cite{chen2021gigaspeech}, and 24K hours of VoxPopuli~\cite{wang2021voxpopuli}.

Previous studies have shown that later transformer layers in WavLM carry more linguistic content, while early layers are likely to encode paralinguistic information~\cite{pasad2023comparative}. For diagnostics, it remains unclear which layers are more crucial. Hence, we aggregated outputs from all 12 layers (with the first input layer excluded) by assigning weights to each of them. These weights were learned through supervised training on downstream tasks. The layer-weighted output from WavLM encoder is a time by feature representation $\{\mathbf{T}\times \mathbf{F}\}$, which can be seen as a temporal representation showing how each feature changes over time. Given the temporal pooling configurations of the CNN layers, the resultant temporal representation has a temporal resolution of \SI{50}{Hz}. However, speech production is modulated at a lower rate and the temporal representation may carry redundant linguistic information that is less essential for disease diagnosis. Thus, we proposed the modulation dynamics block to provide complementary information that is missing from the temporal representation.

\subsection{Modulation dynamics block}
\begin{figure}
    \centering
    \includegraphics[width=\linewidth]{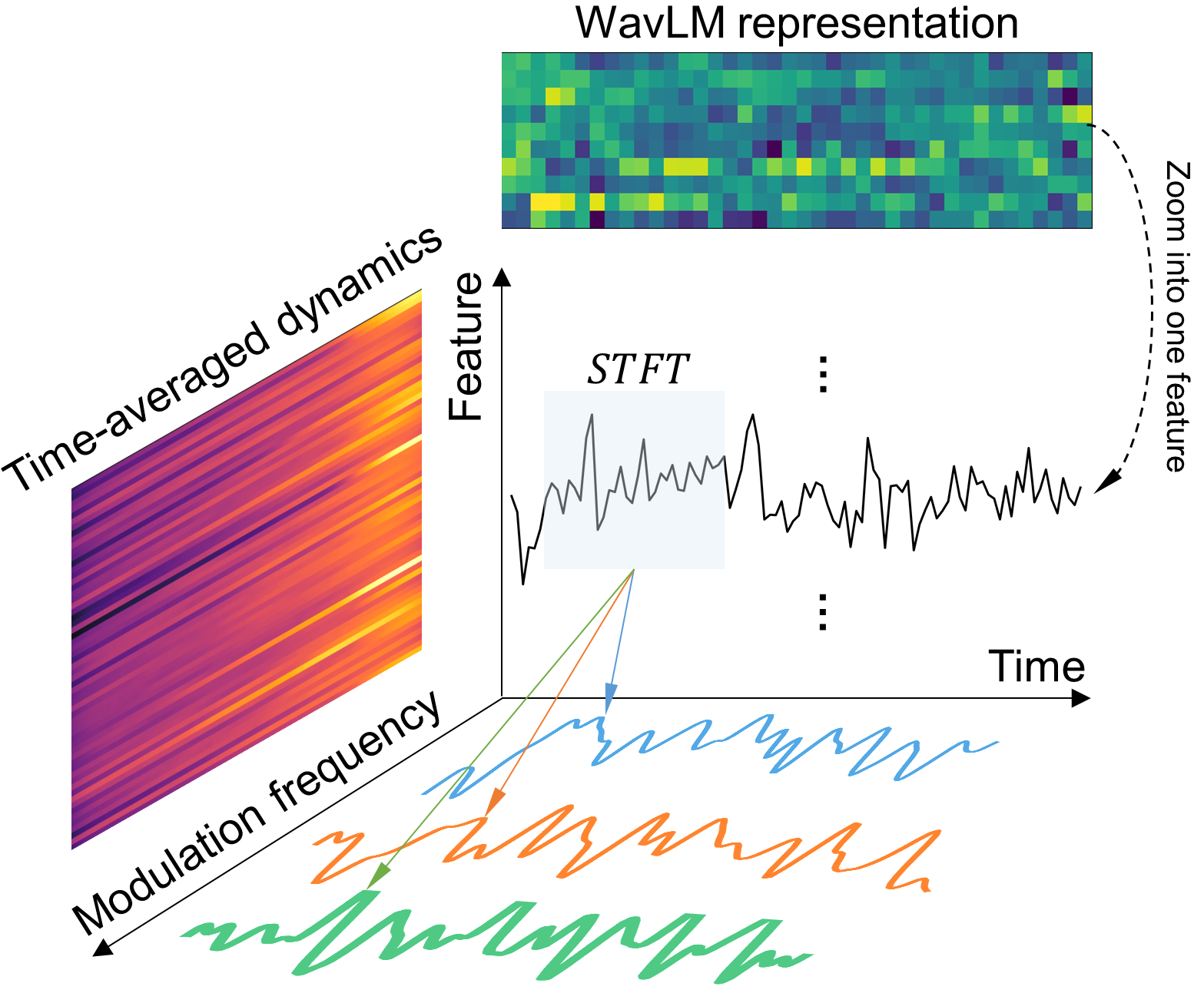}
    \caption{The modulation dynamics block takes the weighted sum of hidden states from the WavLM transformer backbone and applies STFT to each feature channel. }
    \label{fig:mod_block}
\end{figure}

A visual demonstration of the modulation dynamics block is provided in Fig.~\ref{fig:mod_block}. Given an output $\mathbf{T(m,n)}$ from WavLM (i.e., the weight sum of twelve transformer layer outputs), where $\mathbf{m}$ represents the number of time windows and $\mathbf{n}$ represents the number of features, we applied a short-time Fourier transform (STFT) to each feature channel, leading to a 3-dimensional modulation dynamics representation $\mathbf{D_n(j,f_j)}$:
\begin{equation}
    \mathbf{D_n(j,f_j)} = \left|\mathcal{STFT}\mathbf{(T(m,n))}\right|^{2},
\end{equation}
where $\mathbf{j}$ refers to the number of time frames used for STFT and $\mathbf{f_j}$ to the number of modulation frequency channels. The results of the STFT include both real and imaginary parts; here, we keep only the real part by taking the absolute value operation (denoted by $|\cdot|$) and calculate the power. 

For phoneme-level speech applications (e.g., speech recognition), the STFT usually relies on short time windows (e.g.,\SI{16}{}-\SI{32}{ms})~\cite{rabiner2007introduction}, enabling the temporal resolution high enough to discriminate transitory events. The articulatory movement and respiration, on the other hand, are relatively steady and change at a much lower rate than the vibration of vocal cords. Therefore, we extended the window length to $\geq$\SI{128}{ms} with a hop length $\geq$\SI{32}{ms} to capture the dynamics at a wider range. To achieve the optimal performance, we experimented window length values from \SI{128}{ms} to \SI{1}{s} with 25\% hop length, and found the best to be around \SI{256}{ms}. The effects of window sizes are detailed in Section V.C. The resultant dynamics has three different axes, namely feature, time, and modulation frequency, where each slice along the time axis carries the decomposed modulation frequency values for all features.

\subsection{Downstream components}
Similar to speaker embeddings, we assume that the health embeddings correspond to utterance-level characteristics. Thus, both temporal and dynamics representations require a temporal pooling operation to obtain the time-invariant embeddings. We compared average pooling and attentive statistic pooling (ASP) and found the latter to be a better suited for the task at hand. The original ASP aims to integrate the frame-level attention when calculating mean and standard deviation as follows:
\begin{equation}
    \mu = \sum_{t}^{T}\alpha_{t}h_{t},
\end{equation}
\begin{equation}
    \sigma = \sqrt{\sum_{t}^{T}\alpha_{t}h_{t}\odot h_{t}-\mu \odot \mu},
\end{equation}
where $\alpha_t$ represents the weight assigned to the $t$th time frame. 

The ASP can be used directly on temporal features to flatten them into a 1-dimensional vector. With modulation dynamics, we first computed the average along the time axis, which leads to the shape $\{\mathbf{Freq}\times \mathbf{F}\}$, where $\mathbf{Freq}$ stands for frequency and $\mathbf{F}$ for features. We then applied attention to different frequency channels, and calculated the attentive mean and standard deviation. The temporal and dynamics vectors were firstly concatenated then fed into a fully-connected (FC) layer to map into a 768-dimension vector, which was used as the health embedding. A dropout layer and a LeakyReLU with the negative slope of 0.1 were appended after. The second FC layer maps the health embeddings to a single value as the final decision. Additionally, we applied pruning on top of the last FC layer, where the percentage of neurons to be pruned was set as a hyperparameter.

\section{Experimental setup} \label{section:setup}
\subsection{Dataset}
To diversify the types of speech pathologies to be tested, we used six publicly available datasets covering four different speech-related abnormalities. Since they all differ in data collection procedures and some were originally designed for other purposes (e.g., ASR), we here outline the details for each dataset regarding the data collection procedure, data composition and demographics, and data partitions for our downstream ML tasks.
\subsubsection{Respiratory Symptoms Datasets} 
Respiratory symptoms refer to the symptoms induced by infections in the respiratory system, such as coughs, fever, sore throat, etc.~\cite{Resp}. The appearance of respiratory symptoms is commonly seen with asthma, obstructive pulmonary disease (COPD), and pneumonia, just to name a few. At the time of writing, the largest publicly available speech database with various respiratory symptoms is the COVID-19 Sounds~\cite{xia2021covid}. It contains a total of \SI{552}{hours} of audio data recorded remotely from 36,116 individuals around the globe via an app interface. During data collection, volunteers were prompted to conduct three tasks: (1) scripted speech, where all participants uttered the same sentence -- `I hope my data can help to manage the virus pandemic' -- three times in their mother tongue; (2) voluntary cough for three times; and (3) deep breathing through the mouth for three to five minutes. In addition, they also self-reported their COVID-status along with certain metadata information (e.g., gender, age, pre-existing medication condition, respiratory symptoms). In our study, we used only the speech signals and the metadata. It should also be emphasized that not all participants had conducted a PCR test before recording, hence the COVID-status was in the form of a subjective evaluation (e.g., `I think never had COVID-19') rather than a binary label (i.e., positive vs negative). Such ambiguity in COVID-19 labels motivated us to use this database for respiratory abnormality detection instead of COVID-19 prediction.
    
Although the COVID-19 sounds database is advantageous in its size, it may not be the optimal version to train a diagnostics model considering that multiple factors were not controlled, such as language, sampling rate, or acoustic environment. Hence, we set up two subsets from the original database by screening out several potential confounding factors. The first subset was released along with the original database, which was used as the benchmark data for the respiratory symptom prediction task in the COVID-19 Sounds paper~\cite{xia2021covid}. This subset is henceforth referred to as CS-Res. CS-Res contains English samples from 6,623 individuals with respiratory symptoms (e.g., sore throat, cough, etc.), resulting in a total of \SI{31.3}{h} speech data. The sampling rates varied upon different devices used, with the majority sampled at \SI{44.1}{kHz} (67.4\%) and \SI{16}{kHz} (29.8\%). CS-Res was carefully curated so that the recording quality and class balance were controlled. The second subset is similar to CS-Res (in that only English samples are used) but without controlling for the other factors. This subset is referred to as CS-Res-L, with a total of \SI{123.1}{h} of speech, of which 57.1\% were sampled at \SI{16}{kHz} and 40.4\% at \SI{44.1}{kHz} and the rest (2.5\%) were sampled at \SI{8}{kHz} and \SI{12}{kHz} . For both subsets, participants were labelled into two classes, namely the positive ones who reported at least one respiratory symptom, and the negative ones reporting no symptoms at all. With CS-Res, we followed the official partitions as described in~\cite{xia2021covid}. With CS-Res-L, a customized speaker-independent split was performed with a ratio of 7:1:2 (train:validation:test). Meanwhile, we ensured that the distribution of symptom labels, gender, and age were similar in all three splits.
    
\subsubsection{DiCOVA2 Dataset} This dataset contains speech data used in the Second Diagnosing COVID-19 using Acoustics challenge organized in India~\cite{sharma2022second}. DiCOVA2 collected multi-modal acoustic data (i.e., speech, cough, and breathing) remotely from a total of 965 participants via Android and Web apps. Participants were advised to keep the device \SI{10}{cm} from their mouth during recording. For the speech track, participants did number counting from 1 to 20 in a normal pace in English. The recordings were sampled at \SI{48}{kHz}. Furthermore, participants self-reported their metadata, such as gender, experienced symptoms, and COVID-19 status which was grouped into binary labels (either positive or negative). Since the test labels were not made accessible to the public, we used the validation data as the new test set, and partitioned the original training data into the new training set and validation set (8:2).
    
\subsubsection{TORGO Dataset} This dataset consists of speech recordings and synchronized 3D articulatory features collected from healthy controls and speakers with either cerebral palsy (CP) or amyotrophic lateral sclerosis (ALS), the two most prevalent causes of dysarthria~\cite{rudzicz2012torgo}. TORGO was originally designed to develop ASR models for dysarthric individuals. The publicly available version of TORGO includes 8 individuals with dysarthria and 7 healthy controls. During data collection, all subjects were asked to read English text from the screen. The speech data were recorded from two microphones, one facing the participant at a distance of \SI{61}{cm} with a sampling rate of \SI{22.1}{kHz} while the other is head-mounted with a sampling rate of \SI{44.1}{kHz}. Only the data from the front-facing microphone were employed herein. All subjects conducted four different reading tasks: (1) non-words (e.g., high- and low-pitch vowels); (2) short words (e.g. `yes', `no', `back', etc.); (3) restricted sentence (e.g., ``The quick brown fox jumps over the lazy dog''); (4) unrestricted sentence (e.g., spontaneously describe 30 images from the Webber Photo Cards). We included data from all four tasks in our analysis. As there were no official data partitions, we followed the speaker-independent principle to split all 15 subjects into three sets\footnote[1]{`F' and `M' stand for female and male; `C' stands for healthy controls.}: (1) training set (`FC02',`F03',`F01',`MC04',`MC03',`M02'); (2) validation set (`MC02',`FC01',`M03',`M01'); and (3) test set (`FC03',`F04',`MC01',`M05',`M04'). The average dysarthria severity was made similar for all three sets.
    
\subsubsection{Nemours Dataset} This is a collection of speech recordings from 12 males, 11 with different levels of dysarthria and 1 healthy control~\cite{menendez1996nemours}. Each participant was asked to record 74 nonsense sentences of the form ``The $X$ is $Y$ing the $Z$.” ($X\neq Z$). Sentences were generated by randomly selecting $X$ and $Z$ without replacement from a set of 74 monosyllabic nouns and selecting $Y$ without replacement from a set of 37 disyllabic verbs. All recordings were collected in a small sound dampened room with one table-mounted microphone, and digitized subsequently using a \SI{16}{kHz} sampling rate. Apart from recording sessions, Nemours also included a perception session where 5 listeners tried to identify the words of the nonsense sentences. The average number of correct identifications was calculated per speaker and the Frenchay speaker assessment scores were reported, which reflects the severity of dysarthria. The average assessment score of the dysarthric speakers is 74.68 with a standard deviation of 14.54. We labelled all speakers into two classes, namely the relatively severe individuals with scores lower than 74.68 (6 dysarthria speakers), and the mild ones with scores higher than 74.68 (5 dysarthria speakers plus 1 healthy control). 
    
\subsubsection{NCSC Dataset} This refers to the ``NKI CCRT Speech Corpus”~\cite{clapham2012nki}. NCSC contains speech recordings and perceptual evaluations of 55 speakers (10 female and 45 male), who underwent concomitant chemo-radiation treatment (CCRT) for cancer of the head and neck region. Recordings and evaluations were made at three moments: (1) before CCRT; (2) 10-weeks after CCRT; and (3) 12-months after CCRT. All subjects read a 189-word passage from a Dutch fairy tale in a sound-treated room. Speech data were collected using a microphone with a \SI{44.1}{kHz} sampling rate at a distance of \SI{30}{cm} from mouth. 13 speech pathologists rated the intelligibility of these speech recordings on a scale of 1 to 7. We employed the NCSC data released by the \textit{INTERSPEECH 2012 Pathology Sub-Challenge}~\cite{schuller2012interspeech}, where all recordings were labelled either as `intelligible' or `non-intelligible', and were split into three independent sets for model training and evaluation. However, since the test labels were not accessible to the public, we used the validation set as the new test set and split the original training set into the new training and validation set with a ratio of 8:2.

An overview of the data set-up can be found in Table~\ref{tab:dataset}. For reproducibility, we also report if the data split was official or customized, and if a baseline model was released together with the dataset. We further release all data partition details in our code repository for future comparisons. Note that issues were seen with a few samples during our exploratory analysis, such as an empty recording or failures during loading. The file names of these recordings can be found here\footnote[3]{\path{TORGO/FC01/Session1/wav_arrayMic/0256.wav} is an empty recording; Failed to load \path{Nemours/RL/WAV/JPRL39.WAV} with torchaduio.}. These error files were discarded in our experiments.

\renewcommand{\tabcolsep}{5pt}
\begin{table*}
\caption{Employed pathological speech datasets. For reproducibility, we also report if the data split was official and if a baseline model was released together with the dataset, which are indicated by the `Official' and `Baseline' columns. Pos/Neg represents the positive to negative ratio.}
    \centering
    \small
    \begin{tabularx}{\linewidth}{ccccccccccccc}
    \toprule
    \multirow{2}{*}{Pathology} & \multirow{2}{*}{Dataset}& \multirow{2}{*}{Lang} & \multirow{2}{*}{\#hours}  & \multirow{2}{*}{\#spk} & \multirow{2}{*}{\#utt} & \multirow{2}{*}{ave\_dur (\SI{}{s})} & \multicolumn{5}{c}{Data split} & \multirow{2}{*}{Baseline}\\
    \cmidrule(lr){8-12}
    & & & & & & & Official & Pos/Neg & Train & Valid & Test & \\
    \midrule \midrule
    \multirow{2}{*}{Resp symptom} & CS-Res & EN &  31.3 & 6,623 & 9,456 & 11.93$\pm$4.66 & \cmark & 1.05 & 6,648 & 1,914 & 894 & \cmark\\
    & CS-Res-L & EN & 123.1 & 24,134 & 37,140 & 11.94$\pm$4.97& \xmark & 0.78 & 22,308 & 7,969 & 3,863 & \xmark\\
    \midrule
    COVID-19& DiCOVA2 & EN & 3.93 & 975 & 975 & 14.33$\pm$4.15 & \cmark & 0.20 & 617 & 154 & 193 & \cmark\\
    \midrule
    \multirow{2}{*}{Dysarthria} & TORGO & EN & 8.1 & 15 & 9,417 & 3.09$\pm$2.13 & \xmark & 0.51 & 4,564 & 1,753 & 3,100 & \xmark\\
    & Nemours & EN & 1.5 & 12 & 1,628 & 3.35$\pm$ 2.79 & \xmark & 0.38 & 1,184 & 148 & 296 & \xmark\\
    \midrule
    Chemo treated & NCSC & NL & 1.4 & 55 & 1647 & 3.13$\pm$ 1.74 & \cmark & 1.27 & 701 & 200 & 746 & \cmark\\
    \bottomrule
    \end{tabularx}
    \label{tab:dataset}
\end{table*}

\subsection{Benchmark models}
As mentioned in Section~\ref{section:setup}, some challenge datasets were released with a baseline model, namely mel-spectrogram+VGG16 for CS-Res~\cite{xia2021covid}, mel-spectrogram deltas+BiLSTM for DiCOVA2~\cite{sharma2022second}, and openSMILE+RandomForest for NCSC~\cite{schuller2012interspeech}. Though performing well on one dataset, studies have shown that these models lack generalizability across datasets, even within the same type of disease~\cite{zhu2022generalizable}. For simplicity, we group the best performance reported by these baseline models in one row (bottom row in Table~\ref{tab:mp}). Recent work has reported better performance achieved with larger speech models, such as TDNN and transformer-based ones~\cite{desplanques2020ecapa, coppock2022audio}. 

In our study, we compared WavRx to five state-of-the-art speech classification baselines, namely two that leverage SSL encoders, including Wav2vec~\cite{baevski2020wav2vec} and Hubert~\cite{hsu2021hubert}, two different versions of AST pre-trained with speech and audio data respectively~\cite{gong2021ast} (denoted as AST\textsubscript{speech} and AST\textsubscript{audio}), and ECAPA-TDNN~\cite{desplanques2020ecapa}. Modifications were made to these baseline models for compatibility with our tasks. The same ASP layer and classification head implemented in WavRx were appended to Wav2vec and Hubert encoders, and a single FC layer was applied to ECAPA-TDNN embeddings to map these pre-trained representations to a binary output. AST\textsubscript{speech} and AST\textsubscript{audio} were already compatible with our tasks, hence no modifications were made. Three versions of WavRx were compared, namely the original version fusing temporal and dynamics information, and two simplified versions removing either one of the two branches. Details about the baseline models can be found in their corresponding references.


\subsection{Tasks}
To test cross-disease, cross-dataset, and privacy-preserving properties of the proposed method, three tasks are proposed. A fourth task is also included to enhance interpretability. These four tasks include:
\subsubsection{Task 1 -- In-domain diagnostic} This task aims to compare the proposed WavRx to the other baseline models in an in-domain setting. Models were trained and evaluated within each of the six datasets. An ablation study is also conducted to demonstrate the effects of different model components of WavRx. 
\subsubsection{Task 2 -- Zero-shot diagnostic} This task investigates the model generalizability in a stringent setting, where models were trained on one dataset and made predictions on unseen datasets. During inference, both the health embedding encoder and classification head were fixed. This task emulates a scenario where no training data is available from the target domain (e.g., an unseen disease).
\subsubsection{Task 3 -- Privacy of health embeddings} This task examines if the speaker identity is concealed in the WavRx health embeddings by running an automatic speaker verification (ASV) task on top. Since ASV requires multiple recordings from each single individual, TORGO (15 speakers) and Nemours (10 speakers) were selected for this task. With each individual, 10\% of the speech samples were used for training and the remaining 90\% were used for testing. We first extracted the health embeddings using the pre-trained WavRx from Task 1, then applied LDA as the speaker classifier. The WavLM model fine-tuned on Voxceleb 1\&2~\cite{nagrani2017voxceleb} was used as the baseline speaker embedding encoder for comparison purposes.
\subsubsection{Task 4 -- Analysis/interpretability of the modulation dynamics block} Previous tasks have quantified the changes in diagnostic performance, generalizability, and speaker privacy when integrating the modulation dynamics block. This task aims to explore the reason behind these changes by analyzing the characteristics of modulation dynamics and how it shaped the information learned by the upstream WavLM encoder.

\subsection{Training and evaluation details}
For training efficiency, we limited all input recordings to be within \SI{10}{s} by cutting off the over-length part. For those with left and right channels, we took the average to obtain a single-channel audio. All recordings were re-sampled to \SI{16}{kHz} and the amplitude was normalized between -1 and 1. Since the STFT operation in the modulation block requires a minimum of \SI{1}{\second}-signal, short audios were zero-padded to \SI{1}{\second}. The aforementioned pre-processing was achieved using the Torchaudio library \cite{yang2021torchaudio}.

Regarding data augmentation, we injected two types of environmental corruptions in each training batch, namely noise and reverberation, and concatenated the augmented samples with the original samples. Furthermore, we added speed perturbations by slightly speeding up (105\%) and down (95\%) the signal. These approaches were implemented via the SpeechBrain toolkit \cite{speechbrain}.

We used the same hyperparameters for training WavRx on all six datasets, changing only the data augmentation and pruning parameters. These hyperparameters are reported in Table~\ref{tab:hp}. Data augmentation was only used when trained on DiCOVA2 and TORGO; the optimal pruning percentage was set to 90\% for DiCOVA2 and NCSC, and 0\% for the others. With the baseline models, we employed the same data augmentation methods used to train WavRx, and tuned the hyperparameters separately for each one of them.

Diagnostic performance is measured by two metrics, namely the area under the receiver operating curve (AUC-ROC) and the F1 score. The former has been used widely in disease detection tasks as a baseline metric~\cite{sharma2022second, xia2021covid}. However, AUC-ROC has been shown to be over-optimistic when evaluating on extremely imbalanced datasets~\cite{fernandez2018learning}. F1, on the other side, is more robust in an imbalanced setting. With both metrics, we calculated for each class and took the unweighted mean (i.e., \textit{macro}). This is because positive samples (i.e. symptomatic) are usually the minority class, but the missed prediction of a positive sample is more disastrous than that of a negative sample. Hence, the \textit{macro} average is more suitable than the \textit{weighted} average. Furthermore, we found that a model could perform decently on the test set but poorly on the validation set (or vice versa). As such, we report F1 scores achieved with both test and validation sets, where the difference between these two can indicate the model robustness.

Experiments were conducted on the Compute Canada platform~\cite{baldwin2012compute} with four NVIDIA V100-SXM2 (\SI{32}{GB} RAM per GPU). The training time with WavRx was approximately 3-4 hours for CS-Res and CS-Res-L, and less than 2 hours for the other datasets (excluding job waiting time). The shell scripts are also provided in our code repository for simpler replication.

\begin{table}
\centering
\begin{threeparttable} 
\caption{Optimal hyperparameters set for WavRx.}
\label{tab:hp}
    \centering
    \begin{tabularx}{\linewidth}{C|CC}
    \toprule
    Category & Hyper-parameter & Adopted value \\
    \midrule \midrule
    \multirow{9}{*}{Training} & Batch size & 1 \\
    & Learning rate scheduler & Linear \\
    & lr\textsubscript{start} & $1e^-4$ \\
    & lr\textsubscript{end} & $1e^-5$ \\
    & Epochs & 30 \\
    & Optimizer & AdamW \\
    & Early-stop & \cmark\\
    & limit\textsubscript{start} & 2 \\
    & limit\textsubscript{stop} & 3 \\
    \midrule
    \multirow{7}{*}{Model} & STFT window size & \SI{256}{ms} \\
    & STFT hope length & \SI{64}{ms} \\
    & N\_fft & 400 \\
    & Window type & Hamming \\
    & Pad type & Zero-padding \\ 
    & FC dropout & 0.25 \\
    & Pruning percentage$^{\ddagger}$ & 90\% \\
    \midrule
    \multirow{6}{*}{Data aug$^{\dagger}$} & Prob\textsubscript{noise} & 1 \\
    & Prob\textsubscript{reverb} & 1\\
    & SNR\textsubscript{min} & \SI{0}{dB} \\
    & SNR\textsubscript{max} & \SI{15}{dB}\\
    & Speed\textsubscript{min} & 0.95 \\
    & Speed\textsubscript{max} & 1.05 \\
    \bottomrule
    \end{tabularx}
    \begin{tablenotes}\footnotesize
\item $^{\ddagger}$Pruning was used with DiCOVA2 and NCSC. $^{\dagger}$Data augmentation was applied when training on DiCOVA2 and TORGO.
\end{tablenotes}
\end{threeparttable}
\end{table}

\section{Results and Discussion}
\subsection{Task 1: In-domain diagnostic performance}
In-domain diagnostics usually indicates the highest performance that can be achieved by each model in an ideal setting, where training and evaluation data share the same distribution. As shown in Table~\ref{tab:mp}, the proposed WavRx obtains the highest test F1 scores in 4 out of 6 datasets, along with the highest average F1 score of 0.744 (combining test and validation) among all models. With the three datasets that were released with official baseline systems (i.e., CS-Res, DiCOVA2, and NCSC), WavRx markedly outperforms the baselines. When using only the modulation dynamics branch for detection, while the overall performance is not competitive as other benchmarks, it is shown to be the top-performer in the Nemours dysarthria detection task. Together, these results suggest that the dynamics of universal representations is crucial for disease detection.

When comparing different model categories, SSL models (i.e., WavRx, Wav2vec, Hubert) in general outperform those pre-trained in a supervised manner (i.e., AST, ECAPA-TDNN), though both did not include pathological speech during the pre-training. This again demonstrates the benefits of SSL pre-training when evaluated on a variety of downstream tasks. Interestingly, as the only backbone that was pre-trained not on speech data, the AST\textsubscript{audio} outperforms its speech version. Since existing speech foundation models are usually trained with only speech data, potential improvement might be achieved when adding audio data to the pre-training stage, such as music and other non-speech acoustic events.

\renewcommand{\tabcolsep}{3.6pt}
\begin{table*}
\caption{Comparison of model performance on six speech diagnostics datasets. Note that only CS-Res, DiCOVA2, and NCSC had official baselines. `ROC' corresponds to AUC-ROC; `F1\textsubscript{$t$}' refers to test F1 score and `F1\textsubscript{$v$}' to validation F1 score. The last 'Ave' column is the average of F1\textsubscript{$t$} and  F1\textsubscript{$v$} across all datasets. For all three metrics, higher values suggest better performance. Highlighted values represent the best performing model (s) for the metric.}
\centering
\small
\begin{tabularx}{\linewidth}{cccccccccccccccccccc}
     \toprule
     \multirow{3}{*}{Model} & \multicolumn{6}{c}{Respiratory abnormality} & \multicolumn{3}{c}{COVID-19} & \multicolumn{6}{c}{Dysarthria} & \multicolumn{3}{c}{Cancer} & \multirow{3}{*}{Ave}\\
     \cmidrule(lr){2-7}
     \cmidrule(lr){8-10}
     \cmidrule(lr){11-16}
     \cmidrule(lr){17-19}
     & \multicolumn{3}{c}{CS-Res} & \multicolumn{3}{c}{CS-Res-L} & \multicolumn{3}{c}{DiCOVA2} & \multicolumn{3}{c}{TORGO} & \multicolumn{3}{c}{Nemours} & \multicolumn{3}{c}{NCSC} & \\
     \cmidrule(lr){2-4}
     \cmidrule(lr){5-7}
     \cmidrule(lr){8-10}
     \cmidrule(lr){11-13}
     \cmidrule(lr){14-16}
     \cmidrule(lr){17-19}
     & ROC & F1\textsubscript{t} & F1\textsubscript{v} & ROC & F1\textsubscript{t} & F1\textsubscript{v} & ROC & F1\textsubscript{t} & F1\textsubscript{v} & ROC & F1\textsubscript{t} & F1\textsubscript{v} & ROC & F1\textsubscript{t} & F1\textsubscript{v} & ROC & F1\textsubscript{t} & F1\textsubscript{v} & \\
     \midrule
     WavRx & \cellcolor{lightgray}.815 & \cellcolor{lightgray}.730 & \cellcolor{lightgray}.725 & \cellcolor{lightgray}.694  & \cellcolor{lightgray}.655 & \cellcolor{lightgray}.624 & \cellcolor{lightgray}.878 & \cellcolor{lightgray}.600 & .524 & .918 & .767 & \cellcolor{lightgray}.756 & .939 & .959 & .946 & \cellcolor{lightgray}.774 & \cellcolor{lightgray}.737 & .910 & \cellcolor{lightgray}.744\\
     WavRx\textsubscript{mod} & .740 & .645 & .650 & .620 & .571 & .504 & .801 & .550 & .466 & .810 & .659 & .636  & \cellcolor{lightgray}.961 & \cellcolor{lightgray}.980 & .932 & .753 & .716 & .804 & .676\\ 
     WavRx\textsubscript{tem} & .807 & .721 & .720 & .691 & .649 & .594 & .861 & .589 & .478 & .918 & .768 & \cellcolor{lightgray}.756 &  .855 & .872 & .916 & .735 & .695 & .910 & .722\\
     Wav2vec & .798 & .712 & .707 & .682 & .640 & .591 & .841 & .576 & .480 & .827 & .677 & .624 & .945 & .966 & \cellcolor{lightgray}.959 & .759 & .721 & .905 & .713 \\
     Hubert &  .796 & .711 & .707 & .689 & .645 & .592 & .829 & .568 & .479 & \cellcolor{lightgray}.931 & \cellcolor{lightgray}.782 & .687 & .843 & .858 & .973 & .705 & .666 & \cellcolor{lightgray} .928 & .717 \\
     AST\textsubscript{speech} & .683 & .582 & .595 & .610 & .554 &.507 & .738 &.510 &\cellcolor{lightgray}.571 & .762 &.611 &.612 & .727 & .736 & .676 & .636 &.598 &.804 & .613\\
     AST\textsubscript{audio} & .722 & .625 & .661 & .613 &.548 &.584 & .539 &.383 &.462 & .770 & .620 & .603 & .902 & .919 & .703 & .639 &.610 &.822 & .628\\
     ECAPA-TDNN & .687 & .571 & .638 & .582 & .523 & .555 & .755 & .498 & .543 & .636 & .487 & .499 & .640 & .636 & .637 & .703 & .663 & .712 & .580\\
     Baselines & .695 & .594 & .620 & $-$ & $-$  & $-$  & .817 & .561 & .544 & $-$ & $-$ & $-$ & $-$ & $-$ & $-$ & .699 & .658  &  .710 & $-$\\
     \bottomrule
\end{tabularx}
\label{tab:mp}
\end{table*}

\subsection{Task 1: Ablation study}
\begin{figure}
    \centering
    \includegraphics[width=\linewidth]{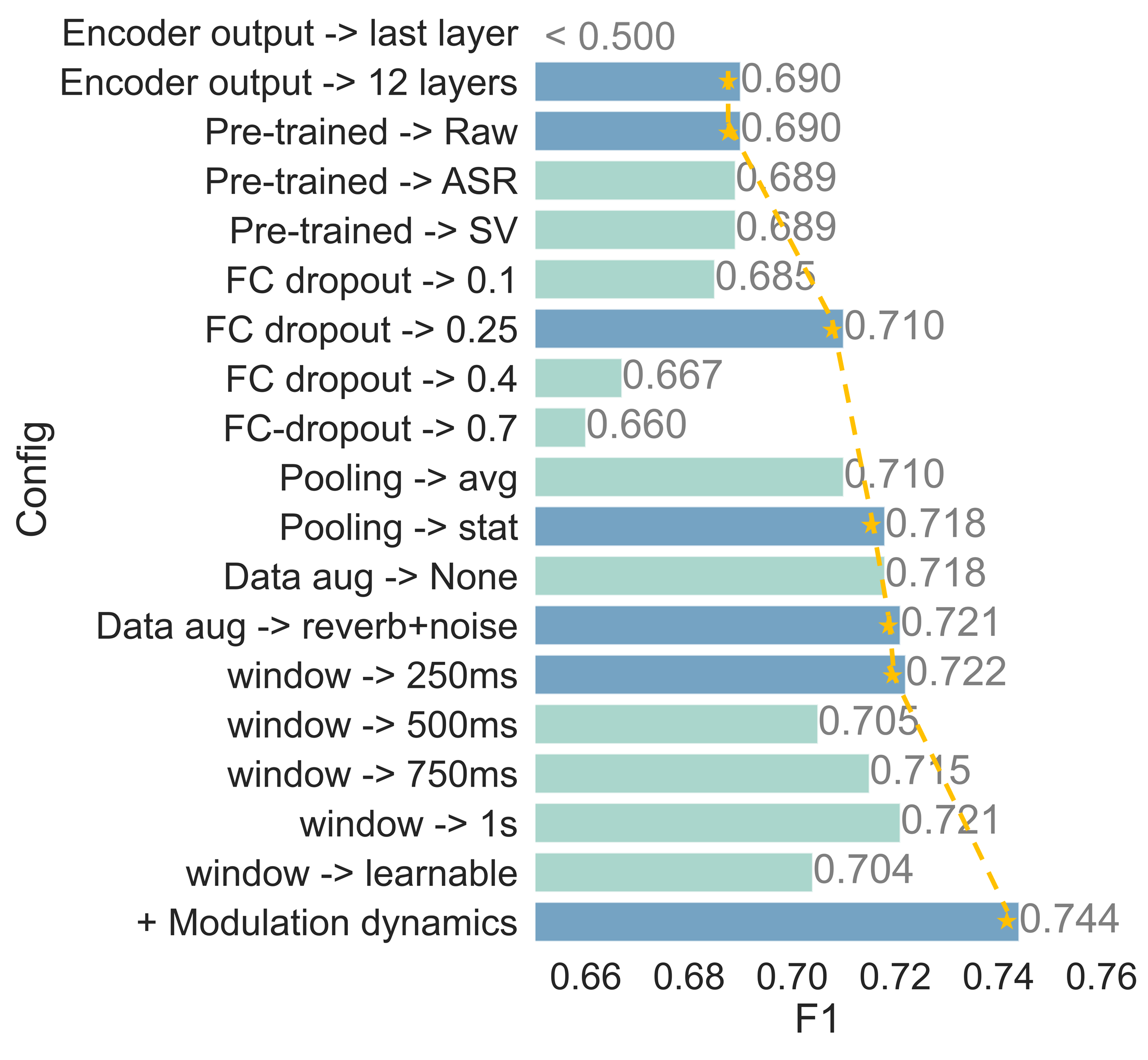}
    \caption{Average F1 scores achieved with different model design choices. The starred ones are the adopted design choices.}
    \label{fig:ablation}
\end{figure}

Next, we carefully examined the improvements brought by the different components of WavRx. Figure~\ref{fig:ablation} shows the improvement in F1 scores averaged across six datasets with different design choices. The largest improvement was seen when all layer outputs were used instead of relying on the last single layer. This corroborates existing SSL model layer analyses, which suggest that later layers likely encode speech semantics while a higher percentage of paralinguistic attributes (e.g., speaker identity, emotion, prosody) are encoded in the early and middle layers~\cite{pasad2021layer,chen2022wavlm, baevski2020wav2vec,ashihara2024self,lin2023utility}. For health diagnostics, the importance of utterance-level characteristics is expected to outweigh frame-level details, since the respiration and articulation patterns do not change from frame to frame. Hence, relying on only the last layer output is suboptimal for health diagnostic tasks. We also explored different WavLM backbone versions, with some further fine-tuned for other tasks, such as ASR and ASV. However, no major difference was seen compared to the raw backbone. The dropout rate is also shown to be important, which helps with model generalization. Since data augmentation is not a major focus of this study, we explored only adding noise and reverberation to the waveforms, which led to minor improvements. Lastly, once other components were optimized, addition of the modulation dynamics branch markedly boosted  overall performance, demonstrating its complementarity to the temporal SSL representation.

\subsection{Task 2: Zero-shot diagnostic performance}
When applied in real-world settings, the amount of data collected from one disease is usually quite limited, as can be seen from the size of the existing pathological speech datasets~\cite{xia2021covid,sharma2022second,menendez1996nemours,schuller2017interspeech,schuller2012interspeech,schuller2021interspeech}. Hence, it can be beneficial when a diagnostic model can generalize to unseen diseases with similar symptoms or pathological origins. As the top-performers in Task 1, we systematically tested WavRx as well as its two individual branches in a cross-dataset setting. Table~\ref{tab:trans} reports the AUC-ROC scores achieved for the model trained on one disease and tested across unseen diseases, as well as the average over all unseen diseases. When comparing different test diseases, respiratory abnormality is shown as the pathology that is distinct from the others, which can be seen from the lowest AUC-ROC score (bottom row in each sub-table). The two dysarthric speech datasets, on the other hand, can lead to decent generalization to each other, although the speech content and data collection protocols differ. Models trained with dysarthric speech can also benefit the detection of COVID-19, as well as chemo-treated speech, which indicates that neuromuscular deficiency can be a shared characteristic among these three pathologies. When comparing the three sub-tables, significant improvements can be seen for all five pathologies when combining modulation dynamics with temporal embeddings. Together with Task 1 results, findings here suggest that integrating modulation dynamics of universal representations can help capture the disease-related biomarkers and improve the model generalizability to diseases sharing similar pathological origins.

\begin{table}
\small
\caption{Cross-disease zero-shot prediction performance using different representations. Values reported are AUC-ROC scores. For each train-test disease combination, the most generalizable representation is color-shaded. Scores without significant difference between the three or below chance-level are ignored. The diagonal values represent the in-domain diagnostic performance.}
    \centering
    \begin{tabularx}{\linewidth}{cccccccc}
    \toprule
    & & \multicolumn{5}{c}{Test set} & \multirow{2}{*}{Ave}\\
    \cmidrule(lr){3-7}
    & & Resp & COVID & Dys-1 & Dys-2 & Cancer & \\
    \midrule
    {\multirow{5}{*}{{\rotatebox[origin=c]{90}{Dyn+Tem}}}} & Resp & \textcolor{lightgray}{.815} & .369 & .489 & \colorbox{lightgray}{.836} & .554 & .613 \\
    & COVID & .493 & \textcolor{lightgray}{.878} & .567 & .693 & .454 & .617 \\
    & Dys-1 & .504 & .684 & \textcolor{lightgray}{.918} & .984 & \colorbox{lightgray}{.690} & .756 \\
    & Dys-2 & .542 & .659 & \colorbox{lightgray}{.763} & \textcolor{lightgray}{.989} & \colorbox{lightgray}{.614} & .713 \\
    & Chemo & .478 & \colorbox{lightgray}{.504} & \colorbox{lightgray}{.638} & \colorbox{lightgray}{.878} & \textcolor{lightgray}{.774} & .654 \\
    \cmidrule(lr){1-7}
    & Ave & \colorbox{lightgray}{.566} & .619 & \colorbox{lightgray}{.675} & \colorbox{lightgray}{.876} & \colorbox{lightgray}{.617} & \\
    \midrule \midrule
    
    {\multirow{5}{*}{{\rotatebox[origin=c]{90}{Dynamics}}}} & Resp & \textcolor{lightgray}{.700} & .447 & \colorbox{pink}{.652} & .708 & \colorbox{pink}{.631} & .628 \\
    & COVID & .498 & \textcolor{lightgray}{.827} & \colorbox{pink}{.635} & \colorbox{pink}{.934} & .346 & .648 \\
    & Dys-1 & .510 & \colorbox{pink}{.759} & \textcolor{lightgray}{.821} & .978 & .631 & .740 \\
    & Dys-2 & .533 & \colorbox{pink}{.798} & .750 & \textcolor{lightgray}{.998} & .495 & .715 \\
    & Chemo & .490 & .337 & .419 & .391 & \textcolor{lightgray}{.753} & .647 \\
    \cmidrule(lr){1-7}
    & Ave & .546 & \colorbox{pink}{.634} & .655 & .802 & .571 & \\
    \midrule \midrule
    
    {\multirow{5}{*}{{\rotatebox[origin=c]{90}{Temporal}}}} & Resp & \textcolor{lightgray}{.721} & \colorbox{cyan}{.598} & .568 & .756 & .552 & .639 \\
    & COVID & .492 & \textcolor{lightgray}{.861} & .580 & .783 & .437 & .631 \\
    & Dys-1 & \colorbox{cyan}{.522} & .600 & \textcolor{lightgray}{.916} & \colorbox{cyan}{.993} & .679 & .742 \\
    & Dys-2 & \colorbox{cyan}{.550} & .406 & .682 & \textcolor{lightgray}{.968} & .563 & .634 \\
    & Chemo & .495 & .378 & .598 & .760 & \textcolor{lightgray}{.746} & .595 \\
    \cmidrule(lr){1-7}
    & Ave & .556 & .569 & .669 & .852 & .595 & \\
    \bottomrule
    \end{tabularx}
    \label{tab:trans}
\end{table}

\subsection{Task 3: Do WavRx health embeddings carry speaker identities?}
Given the system shown in Fig.~\ref{fig:model}, the health embeddings encoded by the local model are expected to carry minimal speaker identity attributes while maximally representing the health information. In this task, we investigate if the health embeddings encoded by the temporal representation alone carry speaker identities, and if the modulation dynamics block can help tackle this issue. The speaker verification accuracies and diagnostic AUC-ROC scores are shown side-by-side in Table~\ref{tab:spk}. Ideally, privacy-preserving health embeddings should have a low ASV accuracy and a high diagnostic AUC-ROC score. As can be seen from row 4 and 7, when relying on only the temporal representation, the learned health embeddings carry a higher amount of speaker identity information than the baseline speaker embeddings. This is likely because that pathological speech follow a different feature distribution than the healthy speech in Voxceleb, hence leading to suboptimal performance of the pre-trained speaker embeddings. The health embeddings obtained from the temporal representation may encode both speaker identity and health attributes, therefore resulting in high ASV accuracies. The modulation dynamics representation, on the other hand, decreases the ASV accuracies by an average rate of 31.9\% and 13.5\% for TORGO and Nemours respectively (row 3 and 6). When fusing the two branches together, the resultant health embeddings lead to the best diagnostic performance, while maintaining the leakage of speaker identity at a lower level than the baseline speaker embeddings (row 2 and 5).

\begin{figure*}
\centering
\includegraphics[width=\linewidth]{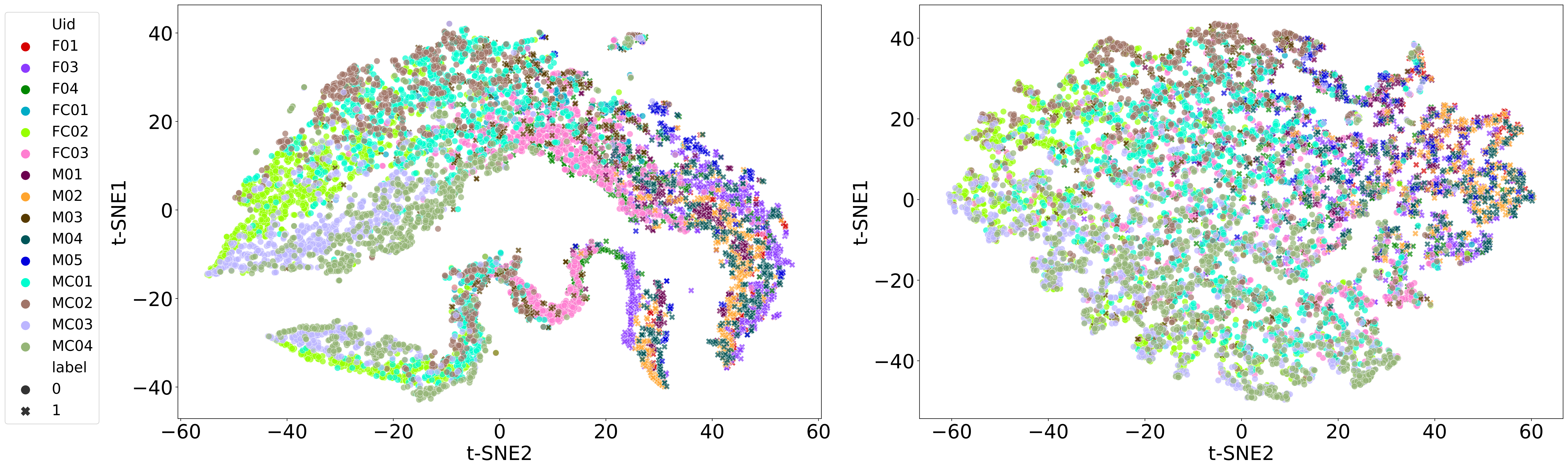}
\caption{Projected health embeddings learned from temporal representations (left) and dynamic representations (right).}
\label{fig:tsne}
\end{figure*}

We further visualize the health embeddings learned from temporal and dynamics representations, which are shown, respectively, in Fig.~\ref{fig:tsne}. Colors represent different speakers and marker types represent disease states. While in both plots, the positive and negative classes can be well separated apart, a more clear distinction between speaker clusters can be seen in the left plot (temporal) than the right one (modulation dynamics), suggesting that speaker identities are better concealed by the proposed modulation dynamics representation. 

\begin{table}
\caption{Speaker verification accuracy and diagnostic AUC-ROC scores obtained by different representations. For ideal health embeddings, we expect lower speaker accuracy and higher diagnostic score.}
\centering
\begin{tabularx}{\linewidth}{ccccc}
\toprule
\multirow{2}{*}{\shortstack[c]{Representation \\ (Model\textsubscript{finetuned dataset})}} & \multicolumn{2}{c}{TORGO} & \multicolumn{2}{c}{Nemours} \\
\cmidrule(l{0.9em}r{0.9em}){2-3} \cmidrule(l{0.9em}r{0.9em}){4-5}
& ACC\textsubscript{spk}$\downarrow$ & AUC\textsubscript{Diag}$\uparrow$ & ACC\textsubscript{spk}$\downarrow$ & AUC\textsubscript{Diag}$\uparrow$ \\
\midrule \midrule
WavLM\textsubscript{Voxceleb} & .715 & - & .951 & -\\
\midrule
WavRx\textsubscript{TORGO} & .711 & \colorbox{lightgray}{.918} & .898 & .984 \\
WavRx-dynamics\textsubscript{TORGO} & .602 & .821 & \colorbox{lightgray}{.831} & .978 \\
WavRx-temporal\textsubscript{TORGO} & .902 & .916 & .990 & .991 \\
\midrule
WavRx\textsubscript{Nemours} & .609 & .763 & .873 & .989 \\
WavRx-dynamics\textsubscript{Nemours} & \colorbox{lightgray}{.594} & .750 & .847 & \colorbox{lightgray}{.998} \\
WavRx-temporal\textsubscript{Nemours} & .857 & .682 & .955 & .967\\
\bottomrule
\end{tabularx}
\label{tab:spk}
\end{table}

\subsection{Task 4: Modulation dynamics analysis and interpretability}
\begin{figure}
    \centering
    \includegraphics[width=\linewidth]{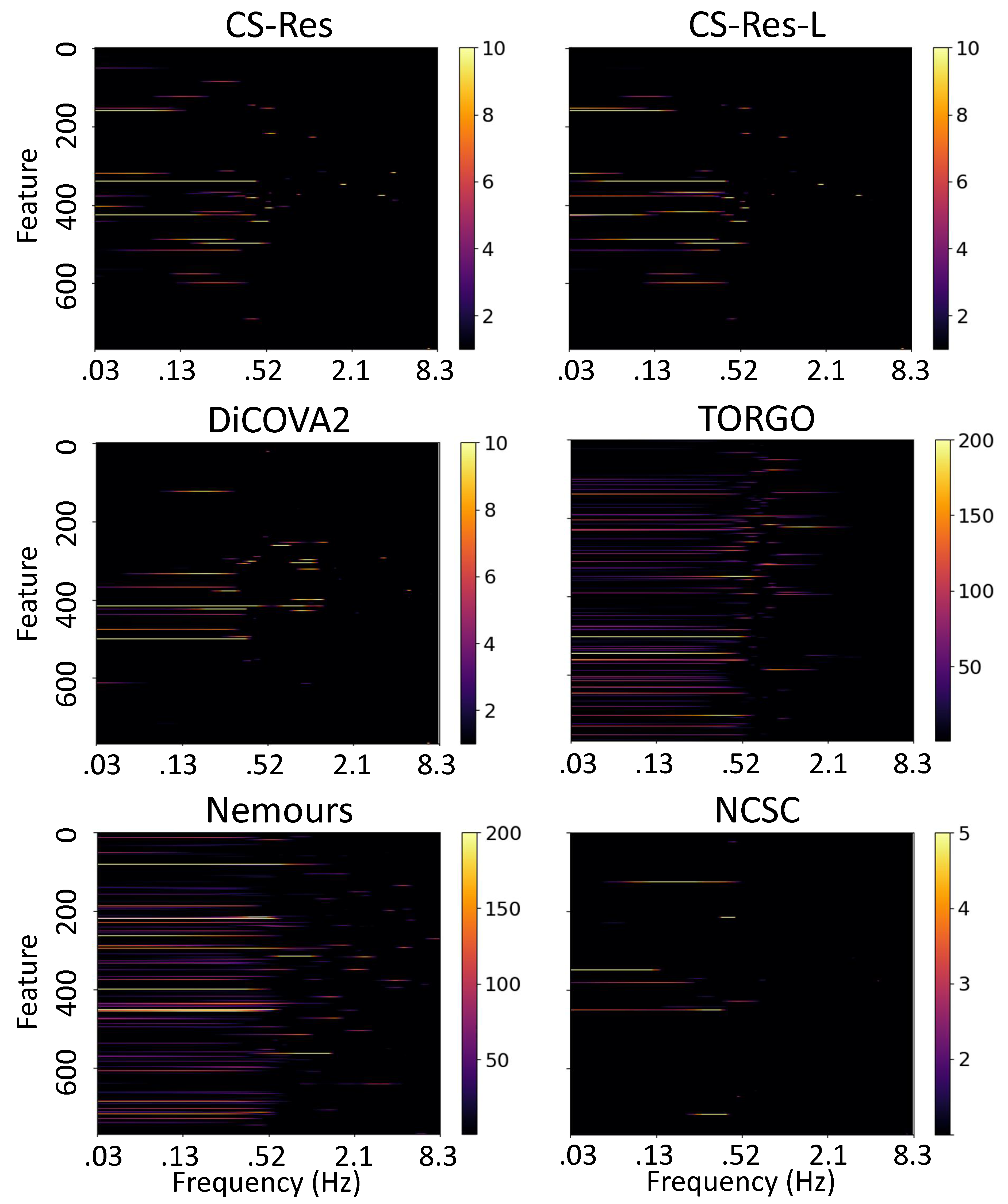}
    \caption{F-ratio plots computed between the modulation dynamics of positive and negative samples obtained for each of the six datasets. X-axis shows the modulation frequency (in \SI{}{\hertz}) and Y-axis represents the feature dimension, which contains 768 features in total. Zoom in on the brighter areas to locate the frequencies, where higher discrimination is obtained between two classes.}
    \label{fig:frs}
\end{figure}

While the modulation dynamics branch is shown to improve the diagnostic performance and generalizability, it is crucial to investigate the characteristics of such representation to understand the reasons behind the improvements. To this end, we start by extracting the modulation dynamic representations from both positive and negative classes, then compute the Fisher's F-ratio~\cite{fisher1970statistical} between the two groups. Since the representation is 2-dimensional (feature by modulation frequency), the F-ratio is calculated per pixel, where the higher value suggests more discrimination between two classes. We further filtered out F-ratio values below 1 since those regions were statistically insignificant. This process was repeated for all six datasets. The F-ratio plots for all tasks can be seen in Fig.~\ref{fig:frs}. 

With the given hop length of the STFT (\SI{64}{\ms}), the maximal modulation frequency is \SI{8.3}{\hertz} with the resolution of \SI{0.125}{\hertz}. For all six datasets, the majority of the difference is observed below \SI{2}{\hertz}, with peaks seen between \SI{0.1}{} - \SI{0.5}{\hertz}, corresponding to a \SI{2} to \SI{5}{\second}-period modulation. Such slow rate of modulation aligns with our initial hypothesis that long-term dynamics of universal representations are crucial for disease detection. While the physiological origin of such modulations still needs to be investigated, it is likely to be associated with slower respiratory and articulatory movement. For example, the automatic contraction of respiratory muscles has been shown to take place once every five seconds during dialogues~\cite{rochet2014take, winkworth1994variability}; an average of 15-25 breathing cycles per minute (equivalent to \SI{2.4}{} to \SI{4}{\second} per cycle) has been reported for adults and the elderly~\cite{barrett2019ganong}.

Another important phenomena noticed is the sparsity of the F-ratio plots, where only very few features among a total of 768 are shown with statistical significance. Based on this observation, we further calculated the sparsity of the 768-dimensional health embeddings learned from dynamic representations and compared with those learned from temporal representations. The sparsity values below 1\% of the per-sample-maximum were thresholded to zeros. The final results are reported in Table~\ref{tab:sparsity}. As can be seen, it is found that the health embeddings learned from temporal representations have an average sparsity of 35.8\% across six datasets with a standard deviation of 9.1\% across samples, while the average sparsity doubles to 76.7\% with only 0.8\% standard deviation for those learned from dynamic representations. Fusing the two together leads to an average sparsity of 64.1\%. Findings here demonstrate that disease-related information is encoded more efficiently by the modulation dynamics, where roughly only half of the features are required for accurately detecting a disease. This not only provides insights into the reasons behind the improved generalizability across diseases, but also helps explain the improved privacy-preserving property of the proposed WavRx model. When learning the health embeddings from the fused representations, health-irrelevant information was likely discarded, which may include speaker attributes, such as gender and age.

\begin{table*}
\centering
\captionsetup{width=.65\linewidth}
\caption{Sparsity of health embeddings learned for each dataset. Sparsity is calculated after thresholding the embedding values.}
\begin{tabularx}{0.65\linewidth}{cccccccc}
\toprule
\multirow{2}{*}{Embedding} & \multicolumn{6}{c}{Sparsity} & \multirow{2}{*}{Average}\\
\cmidrule{2-7}
& Cam-Res & Cam-Res-L & DiCOVA2 & TORGO & Nemours & NCSC & \\
\midrule
Temporal & 33.6$\pm$5.6 & 48.0$\pm$5.0 & 45.7$\pm$10.8 & 28.3$\pm$8.9 & 38.8$\pm$16.3 & 20.6$\pm$8.2 & 35.8$\pm$9.1\\
Dynamics & \bf{88.5$\pm$0.7} & \bf{94.2$\pm$0.1} & \bf{86.6$\pm$0.9} & \bf{65.9$\pm$1.3} & 49.4$\pm$1.2 & \bf{75.4$\pm$0.8} & \bf{76.7$\pm$0.8}\\
Combined & 72.8$\pm$1.7 & 76.6$\pm$1.9 & 64.2$\pm$2.4 & 58.0$\pm$1.4 & \bf{61.2$\pm$1.9} & 52.0$\pm$1.6 & 64.1$\pm$1.8\\
\bottomrule
\end{tabularx}
\label{tab:sparsity}
\end{table*}

\subsection{Task 4: Layer analysis}
\begin{figure}
    \centering
    \includegraphics[width=\linewidth]{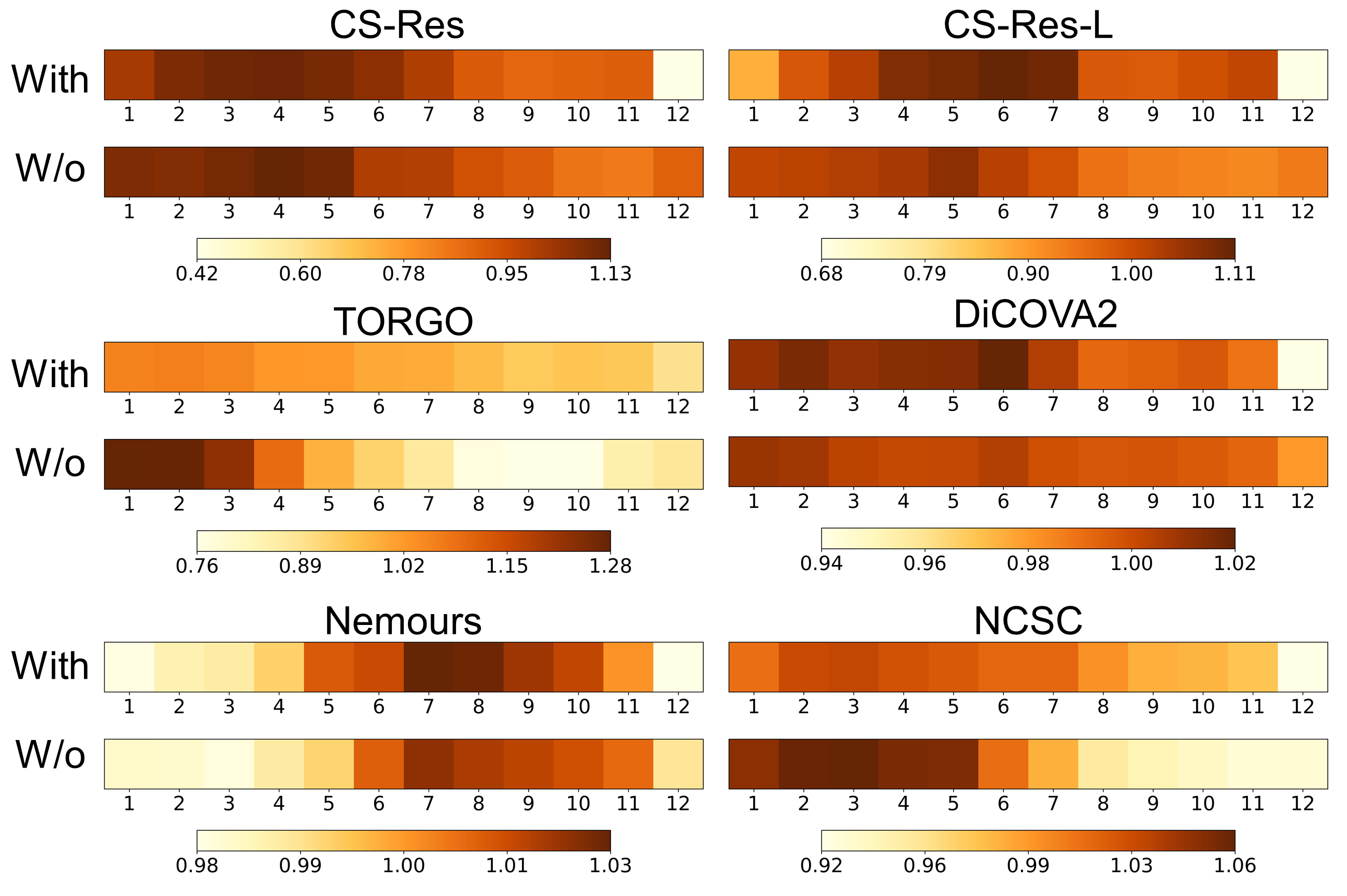}
    \caption{Encoder layer importance learned with and without the proposed modulation dynamics block.}
    \label{fig:layer}
\end{figure}

Similar to a group of studies which performed layer analysis on SSL models for speech applications~\cite{chen2022wavlm, baevski2020wav2vec,ashihara2024self,lin2023utility,pasad2021layer}, we investigated the impact of modulation dynamics block on the learned layer weights. Figure~\ref{fig:layer} compares the layer weights learned with and without the modulation dynamics block. As seen, using only the temporal representation, early layers (0 to 5) are shown to be more crucial, where similar patterns were reported for speaker and emotion recognition tasks~\cite{chen2022wavlm, baevski2020wav2vec,ashihara2024self,lin2023utility,pasad2021layer}. After adding the modulation dynamics, weights are found to shift from early layers to middle layers, with peaks typically seen between layer 6 to 8. Meanwhile, later layers (layer 8 to 11) were also assigned with higher weights. Recent works have suggested that very early layers encode speaker identities~\cite{ashihara2024self}, while middle layers were found most useful for predicting articulation trace~\cite{cho2023evidence}. Combined with our findings, it is likely that modulation dynamics guided the model to focus on articulation-related attributes rather than speaker identities, which led to such shift in layer weights and to the privacy-preserving property observed.

\subsection{Limitations and Future Study}
One potential limitation of our evaluation is the existence of confounding factors in the employed datasets, which have been reported previously~\cite{zhu2023investigating}. Though we have carefully partitioned the datasets so that groups with reported metadata labels are balanced, there might be other underlying factors, such as the noise level, which could lead to over-optimistic results. Furthermore, while the proposed WavRx obtained SOTA performance on majority of the tasks, the robustness to unseen conditions (e.g., in-the-wild speech) can be further improved. This can be seen from the lowest in-domain diagnostic performance achieved with COVID-19 detection, where speech samples were collected in an uncontrolled setting. Meanwhile, recent works have shown the potential of using speech for mental disease detection~\cite{koops2023speech}. While our study did not include such datasets for evaluation, the mechanism of our proposed model could enable its usage for other pathological conditions.

\section{Conclusion}
In this study, we describe a novel speech health diagnostic model termed WavRx, by integrating modulation dynamics with a universal speech representation. Our proposed model achieves SOTA performance on five out of six pathological speech datasets and demonstrate zero-shot generalizability across diseases sharing similar symptoms. Furthermore, the leakage of speaker identities is significantly decreased after adding the innovated modulation dynamics block, thus providing the model with privacy-preserving properties needed in healthcare. An in-depth analysis of the modulation dynamic representation shows that low-frequency modulations below \SI{2}{\hertz} are crucial to discriminate pathological samples. Sparsity and layer analyses help further explain the reasons behind the improvements seen in generalizability and privacy-preserving abilities. In general, WavRx demonstrates generalizability across diseases while minimizing the leakage of speaker identities, hence can be established as a new benchmark model for health diagnostic tasks.


%



\section*{Disclaimer and Acknowledgement}
The authors would like to acknowledge organizations and research groups that made their datasets available to the public. The data holders do not bear any responsibility for the analysis and results presented in this paper. All results and interpretation only represent the view of the authors. Authors also acknowledge funding from INRS, NSERC, and CIHR.

\ifCLASSOPTIONcaptionsoff
  \newpage
\fi


\bibliographystyle{IEEEtran}
\bibliography{ref.bib}

\end{document}